\documentstyle[]{laa}
\def~{\penalty\@M\kern3pt}

\def\@ptsize{0} \@namedef{ds@11pt}{\def\@ptsize{1}}
\@namedef{ds@12pt}{\def\@ptsize{2}}
\def\ds@twoside{\@twosidetrue \@mparswitchtrue}
\def\ds@draft{\overfullrule 5pt}
\newif\if@referee \@refereefalse
\def\ds@referee{\@refereetrue}
\@options

\leftmargin    \parindent
\leftmargini   \leftmargin
\leftmarginii  \leftmargin
\leftmarginiii \leftmargin
\leftmarginiv  \leftmargin
\leftmarginv   \leftmargin
\leftmarginvi  \leftmargin
\labelwidth\leftmargini
\advance\labelwidth-\labelsep
\def\@listii{\leftmargin\leftmarginii
 \labelwidth\leftmarginii\advance\labelwidth-\labelsep
 \topsep \z@}
\def\@listiii{\leftmargin\leftmarginiii
 \labelwidth\leftmarginiii\advance\labelwidth-\labelsep
 \topsep \z@}
\def\@listiv{\leftmargin\leftmarginiv
 \labelwidth\leftmarginiv\advance\labelwidth-\labelsep}
\def\@listv{\leftmargin\leftmarginv
 \labelwidth\leftmarginv\advance\labelwidth-\labelsep}
\def\@listvi{\leftmargin\leftmarginvi
 \labelwidth\leftmarginvi\advance\labelwidth-\labelsep}

\if@referee
\def\@normalsize{\@setsize\normalsize{18dd}\xpt\@xpt
\abovedisplayskip 18pt plus2pt minus5pt
\belowdisplayskip \abovedisplayskip
\abovedisplayshortskip \z@ plus3pt
\belowdisplayshortskip 8pt plus3pt minus3pt
\def\@listi{\topsep 6pt plus 2pt minus 2pt
\leftmargin \leftmargini
\parsep 3pt plus 1pt minus 1pt
\itemsep \parsep}}
\def\small{\@setsize\small{16dd}\ixpt\@ixpt
\abovedisplayskip 15pt plus 3pt minus 4pt
\belowdisplayskip \abovedisplayskip
\abovedisplayshortskip \z@ plus2pt
\belowdisplayshortskip 6pt plus2pt minus 2pt
\def\@listi{\topsep 6pt plus 2pt minus 2pt
\leftmargin \leftmargini
\parsep 3pt plus 1pt minus 1pt
\itemsep \parsep}}
\def\footnotesize{\@setsize\footnotesize{18dd}\xpt\@xpt
\abovedisplayskip 18pt plus 2pt minus 5pt
\belowdisplayskip \abovedisplayskip
\abovedisplayshortskip \z@ plus 1pt
\belowdisplayshortskip 8pt plus 1pt minus 2pt
\def\@listi{\topsep 6pt plus 2pt minus 2pt
\leftmargin \leftmargini
\parsep 3pt plus 1pt minus 1pt
\itemsep \parsep}}
\def\scriptsize{\@setsize\scriptsize{14pt}\viiipt\@viiipt}
\def\tiny{\@setsize\tiny{6pt}\vpt\@vpt}
\def\large{\@setsize\large{24dd}\xivpt\@xivpt}
\def\Large{\@setsize\Large{24dd}\xivpt\@xivpt}
\def\LARGE{\@setsize\LARGE{30dd}\xviipt\@xviipt}
\def\huge{\@setsize\huge{25pt}\xxpt\@xxpt}
\def\Huge{\@setsize\Huge{30pt}\xxvpt\@xxvpt}
\else
\def\@normalsize{\@setsize\normalsize{10dd}\ixpt\@ixpt
\abovedisplayskip 10pt plus2pt minus5pt
\belowdisplayskip \abovedisplayskip
\abovedisplayshortskip \z@ plus3pt
\belowdisplayshortskip 6pt plus3pt minus3pt
\def\@listi{\topsep 4pt plus 2pt minus 2pt
\leftmargin \leftmargini
\parsep 0pt
\itemsep \parsep}}
\def\small{\@setsize\small{9dd}\viiipt\@viiipt
\abovedisplayskip 8.5pt plus 3pt minus 4pt
\belowdisplayskip \abovedisplayskip
\abovedisplayshortskip \z@ plus 2pt
\belowdisplayshortskip 4pt plus2pt minus 2pt
\def\@listi{\topsep 3pt plus 2pt minus 2pt
\leftmargin \leftmargini
\parsep 0pt
\itemsep \parsep}}
\def\footnotesize{\@setsize\footnotesize{10dd}\ixpt\@ixpt
\abovedisplayskip 6pt plus 2pt minus 4pt
\belowdisplayskip \abovedisplayskip
\abovedisplayshortskip \z@ plus 1pt
\belowdisplayshortskip 3pt plus 1pt minus 2pt
\def\@listi{\topsep 4pt plus 2pt minus 2pt
\leftmargin \leftmargini
\parsep 0pt
\itemsep \parsep}}
\def\scriptsize{\@setsize\scriptsize{8pt}\viipt\@viipt}
\def\tiny{\@setsize\tiny{6pt}\vpt\@vpt}
\def\large{\@setsize\large{13dd}\xivpt\@xivpt}
\def\Large{\@setsize\Large{13dd}\xivpt\@xivpt}
\def\LARGE{\@setsize\LARGE{17dd}\xviipt\@xviipt}
\def\huge{\@setsize\huge{25pt}\xxpt\@xxpt}
\def\Huge{\@setsize\Huge{30pt}\xxvpt\@xxvpt}
\fi
\@normalsize

\if@twoside \oddsidemargin 1.5cm \evensidemargin 1.5cm \marginparwidth
107pt \else \oddsidemargin 63pt \evensidemargin 63pt \marginparwidth
90pt \fi
\skip\footins 9pt plus 4pt minus 2pt
-\@lowpenalty \@endparpenalty -\@lowpenalty \@itempenalty -\@lowpenalty

\def\vec#1{\ifmmode\mathchoice{\mbox{\boldmath$\displaystyle#1$}}
{\mbox{\boldmath$\textstyle#1$}}
{\mbox{\boldmath$\scriptstyle#1$}}
{\mbox{\boldmath$\scriptscriptstyle#1$}}\else
\hbox{\boldmath$\textstyle#1$}\fi}

\def\tens#1{\ifmmode\mathchoice{\mbox{$\sf\displaystyle#1$}}
{\mbox{$\sf\textstyle#1$}}
{\mbox{$\sf\scriptstyle#1$}}
{\mbox{$\sf\scriptscriptstyle#1$}}\else
\hbox{$\sf\textstyle#1$}\fi}

\def\part{\par \addvspace{4ex} \@afterindentfalse \secdef\@part\@spart}
\def\@part[#1]#2{\ifnum \c@secnumdepth >\m@ne \refstepcounter{part}
\addcontentsline{toc}{part}{\thepart \hspace{1em}#1}\else
\addcontentsline{toc}{part}{#1}\fi { \parindent 0pt \raggedright \ifnum
\c@secnumdepth >\m@ne \Large \bf Part \thepart \par \nobreak \fi \huge
\bf #2\markboth{}{}\par } \nobreak \vskip 3ex \@afterheading }
\def\@spart#1{{\parindent 0pt \raggedright \huge \bf #1\par} \nobreak
\vskip 3ex \@afterheading }

\def\section{\@startsection {section}{1}{\z@}{-3.5ex plus -1ex minus
 -.2ex}{1.5ex plus .2ex}{\normalsize\bf\boldmath}}
\def\subsection{\@startsection{subsection}{2}{\z@}{-3.25ex plus -1ex
minus -.2ex}{1.5ex plus .2ex}{\normalsize\it}}
\def\subsubsection{\@startsection{subsubsection}{3}{\z@}{-3.25ex plus
-1ex minus -.2ex}{1.5ex plus .2ex}{\normalsize}}
\def\paragraph{\@startsection {paragraph}{4}{\z@}{3.25ex plus 1ex minus
.2ex}{-.6em}{\normalsize\it}}
\def\subparagraph#1{\typeout{AandA Warning: You should not use
\protect\subparagraph \space in this style.}\vskip0.5cm
You should not use $\backslash${\tt subparagraph} in this
style.\vskip0.5cm}

\def\@sect#1#2#3#4#5#6[#7]#8{\ifnum #2>\c@secnumdepth
     \def\@svsec{}\else
     \refstepcounter{#1}\edef\@svsec{\csname the#1\endcsname.\ }\fi
     \@tempskipa #5\relax
      \ifdim \@tempskipa>\z@
 \begingroup #6\relax
   \@hangfrom{\hskip #3\relax\@svsec}{\interlinepenalty \@M #8\par}
 \endgroup
       \csname #1mark\endcsname{#7}\addcontentsline
  {toc}{#1}{\ifnum #2>\c@secnumdepth \else
        \protect\numberline{\csname the#1\endcsname.}\fi
      #7}\else
 \def\@svsechd{#6\hskip #3\@svsec #8\csname #1mark\endcsname
        {#7}\addcontentsline
      {toc}{#1}{\ifnum #2>\c@secnumdepth \else
        \protect\numberline{\csname the#1\endcsname.}\fi
         #7}}\fi
     \@xsect{#5}}

\def\@xsect#1{\@tempskipa #1\relax
      \ifdim \@tempskipa>\z@
       \par \nobreak
       \addvspace{\@tempskipa}
       \@afterheading
    \else \global\@nobreakfalse \global\@noskipsectrue
       \everypar{\if@noskipsec \global\@noskipsecfalse
     \clubpenalty\@M \hskip -\parindent
     \begingroup \@svsechd \endgroup \unskip
     \hskip -#1
    \else       \clubpenalty \@clubpenalty
      \everypar{}\fi}\fi\ignorespaces}

\def\appendix{\par
 \setcounter{section}{0}
 \setcounter{subsection}{0}
 \def\thesection{\Alph{section}}
 \renewcommand{\theequation}{\thesection\arabic{equation}}
 \setcounter{equation}{0}
 \@addtoreset{equation}{section}}

\def\theenumi{\arabic{enumi}}

\def\theenumii{\alph{enumii}}
\def\p@enumii{\theenumi}

\def\theenumiii{\roman{enumiii}}
\def\p@enumiii{\theenumi(\theenumii)}

\def\p@enumiv{\p@enumiii\theenumiii}

\def\verse{\let\\=\@centercr
 \list{}{\itemsep\z@ \itemindent -1.5em\listparindent \itemindent
 \rightmargin\leftmargin\advance\leftmargin 1.5em}\item[]}

\def\descriptionlabel#1{\hspace\labelsep \bf #1}
\def\description{\list{}{\labelwidth\z@ \itemindent-\leftmargin
 \let\makelabel\descriptionlabel}}

\def\theequation{\arabic{equation}}

\def\titlepage{\@restonecolfalse\if@twocolumn\@restonecoltrue\onecolumn
 \else \newpage \fi \thispagestyle{empty}\c@page\z@}
\def\endtitlepage{\if@restonecol\twocolumn \else \newpage \fi}

\tabbingsep \labelsep

\skip\@mpfootins = \skip\footins
\newcounter{part}
\newcounter {section}
\newcounter {subsection}[section]
\newcounter {subsubsection}[subsection]
\newcounter {paragraph}[subsubsection]
\newcounter {subparagraph}[paragraph]

\def\thepart{\Roman{part}} \def\thesection {\arabic{section}}
\def\thesubsection {\thesection.\arabic{subsection}}

\def\@pnumwidth{1.55em}
\def\@tocrmarg {2.55em}
\def\@dotsep{4.5}
\def\tableofcontents{\section*{Contents\markboth{CONTENTS}{CONTENTS}}
 \@starttoc{toc}}
\def\l@part#1#2{\addpenalty{\@secpenalty}
 \addvspace{2.25em plus 1pt} \begingroup
 \@tempdima 3em \parindent \z@ \rightskip \@pnumwidth \parfillskip
-\@pnumwidth
 {\large \bf \leavevmode #1\hfil \hbox to\@pnumwidth{\hss #2}}\par
 \nobreak \endgroup}
\def\l@section#1#2{\addpenalty{\@secpenalty} \addvspace{1.0em plus 1pt}
\@tempdima 1.5em \begingroup
 \parindent \z@ \rightskip \@pnumwidth
 \parfillskip -\@pnumwidth
 \bf \leavevmode #1\hfil \hbox to\@pnumwidth{\hss #2}\par
 \endgroup}
\def\l@subsection{\@dottedtocline{2}{1.5em}{2.3em}}
\def\l@subsubsection{\@dottedtocline{3}{3.8em}{3.2em}}
\def\l@paragraph{\@dottedtocline{4}{7.0em}{4.1em}}
\def\l@subparagraph{\@dottedtocline{5}{10em}{5em}}
\def\listoffigures{\section*{List of Figures\markboth
 {LIST OF FIGURES}{LIST OF FIGURES}}\@starttoc{lof}}
\def\l@figure{\@dottedtocline{1}{1.5em}{2.3em}}
\def\listoftables{\section*{List of Tables\markboth
 {LIST OF TABLES}{LIST OF TABLES}}\@starttoc{lot}}
\let\l@table\l@figure

\def\thebibliography#1{\section*{References\markboth
 {REFERENCES}{REFERENCES}}\list
 {}{\setlength\labelwidth{1.4em}\leftmargin\labelwidth
 \setlength\parsep{0pt}\setlength\itemsep{0pt}
 \setlength{\itemindent}{-\leftmargin}
 \usecounter{enumi}}
 \def\newblock{\hskip .11em plus .33em minus -.07em}
 \sloppy
 \sfcode`\.=1000\relax}

\newif\if@restonecol
\def\theindex{\@restonecoltrue\if@twocolumn\@restonecolfalse\fi
\columnseprule \z@
\columnsep 35pt\twocolumn[\section*{Index}]
 \markboth{INDEX}{INDEX}\thispagestyle{plain}\parindent\z@
 \parskip\z@ plus .3pt\relax\let\item\@idxitem}
\def\@idxitem{\par\hangindent 40pt}

\def\endtheindex{\if@restonecol\onecolumn\else\clearpage\fi}

\def\footnoterule{\kern-3\p@
 \hrule width 2cm
 \kern 2.6\p@}

\long\def\@makefntext#1{\parindent 1em\noindent
 \hbox to 1.8em{\hss$^{\@thefnmark}$}#1}

\def\aafigurecaption{figure}

\long\def\@makecaption#1#2{
 \vskip 10pt
\makeatletter
\ifx\@captype\aafigurecaption\small\fi
\makeatother
 \setbox\@tempboxa\hbox{{\bf #1} #2}
 \ifdim \wd\@tempboxa >\hsize \unhbox\@tempboxa\par \else \hbox
to\hsize{\box\@tempboxa\hfil}
 \fi
\makeatletter
\ifx\@captype\aafigurecaption\else\vskip6pt\fi
\makeatother}

\def\picplace#1{\vbox{\hrule\@height 0.4pt\@width\hsize
\hbox to\hsize{\vrule\@width 0.4pt\@height#1\hfil
\vrule\@width 0.4pt\@height#1}\hrule\@height 0.4pt\@width\hsize}}

\newcounter{figure}
\def\thefigure{\@arabic\c@figure}
\def\fps@figure{htbp}
\def\ftype@figure{1}
\def\ext@figure{lof}
\def\fnum@figure{Fig.\ \thefigure.}
\def\figure{\@float{figure}}
\let\endfigure\end@float
\@namedef{figure*}{\@dblfloat{figure}}
\@namedef{endfigure*}{\end@dblfloat}

\newcounter{table}
\def\thetable{\@arabic\c@table}
\def\fps@table{htbp}
\def\ftype@table{2}
\def\ext@table{lot}
\def\fnum@table{Table \thetable.}
\def\table{\@float{table}}
\let\endtable\end@float
\@namedef{table*}{\@dblfloat{table}}
\@namedef{endtable*}{\end@dblfloat}

\newcounter{@inst}
\newcounter{@auth}
\newdimen\instindent

\def\institute#1{\gdef\@institute{#1}}

\def\institutename{\par
 \begingroup
 \parindent=0pt
 \parskip=0pt
 \setcounter{@inst}{1}%
 \def\and{\par\stepcounter{@inst}%
 \noindent
 \hbox to\instindent{\hss$^{\the@inst}$\enspace}\ignorespaces}%
 \setbox0=\vbox{\def\thanks##1{}\@institute}
 \ifnum\c@@inst>9\relax\setbox0=\hbox{$^{88}$\enspace}%
                 \else\setbox0=\hbox{$^{8}$\enspace}\fi
 \instindent=\wd0\relax
 \ifnum\c@@inst=1\relax\else
 \setcounter{@inst}{1}%
 \hangindent\instindent\hangafter=0\noindent
 \llap{$^{\the@inst}$\enspace}\fi\ignorespaces
 \@institute\par
 \endgroup}

\def\offprints#1{\begingroup
\def\protect{\noexpand\protect\noexpand}\xdef\@thanks{\@thanks
\protect\footnotetext[0]{\unskip{\it Send offprint requests
to\/}: \ignorespaces#1}}\endgroup\ignorespaces}

\def\@thanks{}

\long\def\@makefntext#1{\parindent 1em\noindent
$^{\@thefnmark}$#1}

\def\@fnsymbol#1{\ifcase#1\or\star\or{\star\star}\or{\star\star\star}%
   \or \dagger\or \ddagger\or
   \mathchar "278\or \mathchar "27B\or \|\or **\or \dagger\dagger
   \or \ddagger\ddagger \else\@ctrerr\fi\relax}

\def\subtitle#1{\gdef\@subtitle{#1}}
\def\@subtitle{}

\def\thesaurus#1{\gdef\@thesaurus{#1}}
\def\@thesaurus{missing; you have not inserted them}%

\def\maketitle{\par
 \begingroup
 \def\thefootnote{\fnsymbol{footnote}}
 \if@twocolumn
   \twocolumn[\@maketitle]
 \else
   \newpage \@maketitle
 \fi
 \global\@topnum\z@
 \thispagestyle{empty}\@thanks
 \endgroup
 \setcounter{footnote}{0}
 \let\maketitle\relax
 \let\@maketitle\relax
 \gdef\@thanks{}\gdef\@author{}\gdef\@title{}\gdef\@subtitle{}%
 \let\thanks\relax}

\def\AALogo{\setbox254=\hbox{ ASTROPHYSICS }%
\vbox{\baselineskip=10dd\hrule\hbox{\vrule\vbox{\kern3pt
\hbox to\wd254{\hfil ASTRONOMY\hfil}
\hbox to\wd254{\hfil AND\hfil}\copy254
\hbox to\wd254{\hfil\number\day.\number\month.\number\year\hfil}
\kern3pt}\vrule}\hrule}}
\def\makeheadbox{{\hsize=30cc
\hbox to0pt{\vbox{\baselineskip=10dd\hrule\hbox
to\hsize{\vrule\kern3pt\vbox{\kern3pt
\hbox{\bf A\&A manuscript no.}
\hbox{(will be inserted by hand later)}
\kern3pt\hrule\kern3pt\bf
\hbox{Your thesaurus codes are:}
\hbox{\rightskip=0pt plus3em\advance\hsize by-7pt
\vbox{\noindent\ignorespaces\@thesaurus}}
\kern3pt}\hfil\kern3pt\vrule}\hrule}
\rlap{\quad\AALogo}\hss}}}

\def\@maketitle{\newpage
 \rm\vbox to0pt{}\vskip-8mm
 \makeheadbox
 \vskip13.5mm
 {\LARGE \bf\boldmath
  \pretolerance=10000
  \rightskip=0pt plus 4cm
  \noindent\ignorespaces
  \@title \par}\vskip .3cm
\if!\@subtitle!\else {\Large \bf\boldmath
  \vskip .05cm
  \pretolerance=10000
  \rightskip=0pt plus 3cm
  \noindent\@subtitle \par}\vskip .4cm\fi
 {\bf \lineskip .5em
\setbox0=\vbox{\setcounter{@auth}{1}\def\and{\stepcounter{@auth}}%
\def\thanks##1{}\@author\global\c@@inst=\c@@auth}%
\def\lastand{\ifnum\c@@inst=2\relax\unskip{} and \else
\unskip, and \fi}%
\setcounter{@auth}{1}%
\def\and{\stepcounter{@auth}\ifnum\c@@auth=\c@@inst\lastand\else
\unskip, \fi}%
 \noindent\@author\vskip.125cm}
 \institutename
 \vskip .35cm {\normalsize\noindent\@date}
 \par
 \vskip .7cm}

\mark{{}{}}

\if@twoside
\def\ps@headings{\def\@oddfoot{}\def\@evenfoot{}\def\@evenhead{\rm
\thepage\hfil \sl \leftmark}\def\@oddhead{\hbox{}\sl \rightmark \hfil
\rm\thepage}\def\sectionmark##1{\markboth {\uppercase{\ifnum
\c@secnumdepth >\z@ \thesection\hskip 1em\relax \fi
##1}}{}}\def\subsectionmark##1{\markright {\ifnum \c@secnumdepth >\@ne
\thesubsection\hskip 1em\relax \fi ##1}}}
\else
\def\ps@headings{\def\@oddfoot{}\def\@evenfoot{}\def\@oddhead{\hbox
{}\sl \rightmark \hfil \rm\thepage}\def\sectionmark##1{\markright
{\uppercase{\ifnum \c@secnumdepth >\z@ \thesection\hskip 1em\relax \fi
##1}}}}
\fi
\def\ps@myheadings{\def\@oddhead{\hbox{}\hfil\rm\thepage\hfil
}\def\@oddfoot{}\def\@evenhead{\hfil\rm\thepage\hfil\hbox
{}}\def\@evenfoot{}\def\sectionmark##1{}\def\subsectionmark##1{}}

\def\today{\ifcase\month\or
 January\or February\or March\or April\or May\or June\or
 July\or August\or September\or October\or November\or December\fi
 \space\number\day, \number\year}

\ps@myheadings  \onecolumn
\if@twoside\else\fi

\if@referee
  \onecolumn
\else
  \twocolumn
\fi
\def\@array[#1]#2{\setbox\@arstrutbox=\hbox{\vrule
     height\arraystretch \ht\strutbox
     depth\arraystretch \dp\strutbox
     width\z@}\@mkpream{@{}#2}\edef\@preamble{\halign \noexpand\@halignto
\bgroup \tabskip\z@ \@arstrut \@preamble \tabskip\z@ \cr}%
\let\@startpbox\@@startpbox \let\@endpbox\@@endpbox
  \if #1t\vtop \else \if#1b\vbox \else \vcenter \fi\fi
  \bgroup \let\par\relax
  \let\@sharp##\let\protect\relax \lineskip\z@\baselineskip\z@\@preamble}

\def\[{\relax\ifmmode\@badmath\else\begingroup\trivlist
\item[]\leavevmode \hbox to\linewidth\bgroup$ \displaystyle
\hskip\mathindent\bgroup\fi}
\def\]{\relax\ifmmode \egroup $\hfil \egroup \endtrivlist \endgroup\else
\@badmath \fi}
\def\equation{\refstepcounter{equation}\trivlist \item[]\leavevmode
\hbox to\linewidth\bgroup $ \displaystyle \hskip\mathindent}
\def\endequation{$\hfil \displaywidth\linewidth\@eqnnum\egroup
\endtrivlist}
\def\eqnarray{\stepcounter{equation}\let\@currentlabel=\theequation
\global\@eqnswtrue \global\@eqcnt\z@\tabskip\mathindent\let\\=\@eqncr
\abovedisplayskip\topsep\ifvmode\advance\abovedisplayskip\partopsep\fi
\belowdisplayskip\abovedisplayskip
\belowdisplayshortskip\abovedisplayskip
\abovedisplayshortskip\abovedisplayskip $$\halign to
\linewidth\bgroup\@eqnsel\hskip\@centering$\displaystyle\tabskip\z@
{##}$&\global\@eqcnt\@ne \hskip 2\arraycolsep \hfil${##}$\hfil
&\global\@eqcnt\tw@ \hskip 2\arraycolsep $\displaystyle{##}$\hfil
\tabskip\@centering&\llap{##}\tabskip\z@\cr}
\def\endeqnarray{\@@eqncr\egroup
\global\advance\c@equation\m@ne$$\global\@ignoretrue}
\newdimen\mathindent
\newenvironment{abstract}{\begin{abstr}\ignorespaces}{\vskip0.5cm\hrule
\vskip3ptplus10dd\null\end{abstr}}

\def\keywords{\par\vspace{12pt}\noindent{\bf Key words: }}

\def\@cite#1#2{{#1\if@tempswa , #2\fi}}
\def\@biblabel#1{}

\def\squareforqed{\hbox{\rlap{$\sqcap$}$\sqcup$}}
\def\sq{\ifmmode\squareforqed\else{\unskip\nobreak\hfil
\penalty50\hskip1em\null\nobreak\hfil\squareforqed
\parfillskip=0pt\finalhyphendemerits=0\endgraf}\fi}

\def\utw{\smash{\rlap{\lower5pt\hbox{$\sim$}}}}
\def\udtw{\smash{\rlap{\lower6pt\hbox{$\approx$}}}}

\def\diameter{{\ifmmode\mathchoice
{\ooalign{\hfil\hbox{$\displaystyle/$}\hfil\crcr
{\hbox{$\displaystyle\mathchar"20D$}}}}
{\ooalign{\hfil\hbox{$\textstyle/$}\hfil\crcr
{\hbox{$\textstyle\mathchar"20D$}}}}
{\ooalign{\hfil\hbox{$\scriptstyle/$}\hfil\crcr
{\hbox{$\scriptstyle\mathchar"20D$}}}}
{\ooalign{\hfil\hbox{$\scriptscriptstyle/$}\hfil\crcr
{\hbox{$\scriptscriptstyle\mathchar"20D$}}}}
\else{\ooalign{\hfil/\hfil\crcr\mathhexbox20D}}%
\fi}}

\def\bbbc{{\mathchoice {\setbox0=\hbox{$\displaystyle\rm C$}\hbox{\hbox
to0pt{\kern0.4\wd0\vrule height0.9\ht0\hss}\box0}}
{\setbox0=\hbox{$\textstyle\rm C$}\hbox{\hbox
to0pt{\kern0.4\wd0\vrule height0.9\ht0\hss}\box0}}
{\setbox0=\hbox{$\scriptstyle\rm C$}\hbox{\hbox
to0pt{\kern0.4\wd0\vrule height0.9\ht0\hss}\box0}}
{\setbox0=\hbox{$\scriptscriptstyle\rm C$}\hbox{\hbox
to0pt{\kern0.4\wd0\vrule height0.9\ht0\hss}\box0}}}}
\def\bbbq{{\mathchoice {\setbox0=\hbox{$\displaystyle\rm
Q$}\hbox{\raise
0.15\ht0\hbox to0pt{\kern0.4\wd0\vrule height0.8\ht0\hss}\box0}}
{\setbox0=\hbox{$\textstyle\rm Q$}\hbox{\raise
0.15\ht0\hbox to0pt{\kern0.4\wd0\vrule height0.8\ht0\hss}\box0}}
{\setbox0=\hbox{$\scriptstyle\rm Q$}\hbox{\raise
0.15\ht0\hbox to0pt{\kern0.4\wd0\vrule height0.7\ht0\hss}\box0}}
{\setbox0=\hbox{$\scriptscriptstyle\rm Q$}\hbox{\raise
0.15\ht0\hbox to0pt{\kern0.4\wd0\vrule height0.7\ht0\hss}\box0}}}}
\def\bbbt{{\mathchoice {\setbox0=\hbox{$\displaystyle\rm
T$}\hbox{\hbox to0pt{\kern0.3\wd0\vrule height0.9\ht0\hss}\box0}}
{\setbox0=\hbox{$\textstyle\rm T$}\hbox{\hbox
to0pt{\kern0.3\wd0\vrule height0.9\ht0\hss}\box0}}
{\setbox0=\hbox{$\scriptstyle\rm T$}\hbox{\hbox
to0pt{\kern0.3\wd0\vrule height0.9\ht0\hss}\box0}}
{\setbox0=\hbox{$\scriptscriptstyle\rm T$}\hbox{\hbox
to0pt{\kern0.3\wd0\vrule height0.9\ht0\hss}\box0}}}}
\def\bbbs{{\mathchoice
{\setbox0=\hbox{$\displaystyle     \rm S$}\hbox{\raise0.5\ht0\hbox
to0pt{\kern0.35\wd0\vrule height0.45\ht0\hss}\hbox
to0pt{\kern0.55\wd0\vrule height0.5\ht0\hss}\box0}}
{\setbox0=\hbox{$\textstyle        \rm S$}\hbox{\raise0.5\ht0\hbox
to0pt{\kern0.35\wd0\vrule height0.45\ht0\hss}\hbox
to0pt{\kern0.55\wd0\vrule height0.5\ht0\hss}\box0}}
{\setbox0=\hbox{$\scriptstyle      \rm S$}\hbox{\raise0.5\ht0\hbox
to0pt{\kern0.35\wd0\vrule height0.45\ht0\hss}\raise0.05\ht0\hbox
to0pt{\kern0.5\wd0\vrule height0.45\ht0\hss}\box0}}
{\setbox0=\hbox{$\scriptscriptstyle\rm S$}\hbox{\raise0.5\ht0\hbox
to0pt{\kern0.4\wd0\vrule height0.45\ht0\hss}\raise0.05\ht0\hbox
to0pt{\kern0.55\wd0\vrule height0.45\ht0\hss}\box0}}}}
\def\bbbz{{\mathchoice {\hbox{$\sf\textstyle Z\kern-0.4em Z$}}
{\hbox{$\sf\textstyle Z\kern-0.4em Z$}}
{\hbox{$\sf\scriptstyle Z\kern-0.3em Z$}}
{\hbox{$\sf\scriptscriptstyle Z\kern-0.2em Z$}}}}

\def\typeset{\vfill{\small\noindent This article was processed by the
author using Springer-Verlag \LaTeX\ A\&A style file 1990.\par}}

\def\enddocument{\par\typeset
\@checkend{document}\clearpage\begingroup
\if@filesw \immediate\closeout\@mainaux
\def\global\@namedef##1##2{}\def\newlabel{\@testdef r}%
\def\bibcite{\@testdef b}\@tempswafalse\makeatletter\input \jobname.aux
\if@tempswa \@warning{Label(s) may have changed. Rerun to get
cross-references right}\fi\fi\endgroup\deadcycles\z@\@@end}

\newcommand{\be}{\begin{equation}}
\newcommand{\ee}{\end{equation}}

\def\lsim{~\rlap{$<$}{\lower 1.0ex \hbox{$\sim$}}}
\def\gsim{~\rlap{$>$}{\lower 1.0ex \hbox{$\sim$}}}
\newcommand{\kmsec}{\,\mbox{$\mbox{km}\,\mbox{s}^{-1}$}}

\begin{document}
%
\thesaurus{06(08.01.2; 08.03.3; 08.12.1; 08.13.1; 08.18.1; 10.15.1)}
\title{New observational limits on dynamo saturation
in young solar-type stars}

\author{M.A. O'Dell
  \and  P. Panagi
  \and  M.A. Hendry
  \and  A. Collier Cameron}
\offprints{M.A. O'Dell}
\institute{Department of Astronomy, University
of Sussex, Falmer, Brighton BN1 9QH, UK}
\date{Received date; accepted date}
%
\maketitle
\begin{abstract}
We present statistically robust observational evidence which imposes new
limits on dynamo saturation in young solar-type stars.
These are inferred from the increasing amplitude of the V-band optical
flux with rotation,
caused by the filling of the disc with surface spots in a
non-axisymmetric pattern.

Assuming spot coverage acts as a tracer of the total magnetic surface
flux we find
that the magnetic activity saturates at a level at least 6 -- 10 times greater
than that inferred from chromospheric and transition line indicators.

We suggest that these new limits imply a minimum rotation
for saturation of the dynamo and that for high rotation rates starspot
coverage acts
as an alternative diagnostic for the stellar dynamo to the chromospheric
and transition region line emission fluxes.

The fact that the dynamo does not appear to
saturate at the low rotation rates
indicated by chromospheric indicators
should assist evolutionary braking models that have to explain the
sudden spin-down of young fast rotating G-dwarfs.

\end{abstract}

\section{Introduction}
The magnetic fields observed in late-type stars are believed to be a
consequence of a dynamo action that resides close to the interface of
the stars' radiative core and convective envelope (Parker 1955).
This dynamo action operates by virtue of the star's rotation on its
axis. Saar (1991), hereafter S91, observed that the magnetic
flux density at the
stellar surface essentially increases linearly with rotation
($fB \, \propto \, {\Omega}^{\alpha}$, $\alpha \sim1$) where $f$
(the fractional filling parameter) rather
than $B$ (the mean field strength) is the determining factor.
With increasing rotation, active regions are thought to spread and fill
the stellar surface ($f \rightarrow 1$) and surface magnetic activity
is believed to
saturate. Evidence of the saturation of magnetic activity
from chromospheric UV and
optical line flux
indicators has been shown by Vilhu 1984 (hereafter V84). Further
evidence of saturation has been inferred from soft X-ray
observations of the transition region
(Vilhu and Rucinski 1983) and unpolarized Zeeman broadening of
photospheric line profiles (S91).

It has long been realised that the chromospheric activity observed in
late-type stars is related to rotation rate (e.g. Kraft 1967, Skumanich
1972). More recently Noyes et al (1984), hereafter N84,
confirmed a correlation between
Ca II H+K emission and axial rotation period and found a further
dependence of the emission on spectral type.
Chromospheric and transition region emission line fluxes are often employed as
proxy indicators of magnetic activity but are reliant on the assumption that
nonthermal chromospheric heating processes relate the surface field to
the emission (N84). Indeed magnetic heating processes are most likely
responsible for the observed chromospheric emission in stars but the
contribution of acoustic heating by waves generated in the convection
zone is not clear (e.g. review by Linsky 1990).
It does appear, however, that the chromospheric and transition
region fluxes saturate at an axial rotation period ( $\sim$3 days for
solar-type stars, Vilhu and Rucinski (1983), V84),
which corresponds to about ten times the solar value.
This was confirmed by S91 from unpolarised Zeeman broadened line profiles,
who
observed a saturation of the filling parameter $f$ at roughly the same
value of the activity parameter $R = {\tau}_c/P_{rot}$ as was inferred
by V84, where ${\tau}_c$ is the convective
turnover time and $P_{rot}$ is the axial rotation period.
This is to be expected as Saar's technique is biased
towards the bright photospheric network which is associated with
active regions such as plages. Because of their low surface
brightnesses, starspots are expected
to contribute little to the line profiles (Saar
1988).

In this paper we consider the amplitude of photometric
variability as an indicator of the spot coverage, and hence
of total magnetic surface flux, for solar-type stars.
Variations in the broadband photometric light curves were first
suggested to be caused by surface inhomogeneities such as starspots
by Kron (1952).
The purpose of this paper is twofold: to examine whether
${\Delta}V$ ceases to increase beyond some critical rotation
rate ($\tilde{\Omega}$) (therefore
implying maximum spot coverage and saturation of the dynamo) and
to compare this result with that inferred from
chromospheric indicators.

In order to understand how ${\Delta}V$ varies with rotation, an
homogeneous sample of solar-type stars was chosen whose axial
rotation periods and light curve amplitudes were known. The sources of
these data are given in Table 1 of section 2.

In section 3 we investigate how ${\Delta}V$ correlates with the
logarithm of
the activity parameter $R$, and make direct comparisons with
the work of V84 and S91.
We infer a lower limit on the value of the rotation rate (or activity
parameter) at which the
photometric amplitude or spot coverage saturates, as compared to
that indicated by  chromospheric and transition line indicators.
We also discuss the possible dependence of ${\Delta}V$ on ($B-V$)
colour.

Finally in section 4 we discuss the usefulness of the photometric
amplitude and
chromospheric emission lines fluxes as proxies for the dynamo-generated
magnetic flux threading the stellar surface. We also discuss the
possible implications for magnetic braking theory.

\section{The observational data}\label{sec:obsdata}

Since the discovery by van Leeuwen and Alphenaar (1982) that a number of
the young late-type members of the Pleiades cluster exhibit high
rotation rates, $v{\sin}i \sim{100}\kmsec$, enormous effort has been
made via differential photometry to secure the
rotational periods of other young cluster members. This has resulted
in numerous light-curves and hence short term photometric
amplitudes.
We therefore draw from this pool of data, including in our
sample members of the Pleiades and $\alpha$
Persei open clusters, (ages: 70Myr and 50Myr respectively (Mermilliod
1981)),
local young G-dwarfs and purported members
of the Pleiades moving group.

As we are considering only solar-type stars we restrict the
spectral range of our sample, maintaining an
homogeneity, by considering only single G and K
dwarfs in the colour range ${0.55}\leq{B-V}\leq{1.40}$. The data
available restricted us to ages in excess of 50Myr. We included young
field dwarfs to age $\sim2$Gyr to extend the long-period end of our
sample.
This age constraint reduced
possible anomalous effects caused by dynamo evolution (Gilliland 1985,
hereafter G85).

A proportion of young late-type stars
arrive on the main-sequence with high rotation rates (Stauffer et al
1985). This results in clusters such as $\alpha$ Persei, and to a
lesser extent the older Pleiades, exhibiting broad distributions
of rotational velocity.
This allowed us to examine the amplitude of the
modulated V-band flux through
a complete range of rotation rates, from twice solar to near
break-up velocity.

Table 1 lists all the stars employed in this study, their respective
photometric and rotational data and lastly their source of reference.

\section{Photometric variability, rotation and saturation}\label{sec:vrs}

When considering surface magnetic activity and its relation to rotation,
Noyes (1983) noticed that the correlation of chromospheric emission
fluxes with rotation was improved scaling an angular rotation rate,
$\Omega_{n}$ ($= {P_{rot}}^{-1}$), by a function of spectral type.
This function, $g(B-V)$, approximates closely to the
variation of the convective turnover time, ${\tau}_c$, with
spectral type, at the base of
the convective zone, computed empirically or
from stellar structure models.
The product of ${\tau}_c$ and $\Omega_{n}$ is known as
the
inverse Rossby number, the activity parameter $R$,
of V84 and S91 viz.
\begin{equation}
R = {\tau_c}{\Omega_{n}} = {\tau_c}/P_{rot}
\end{equation}

V84 observed the chromospheric and transition region line emission
fluxes to saturate when $R{\approx}3$, which he
interpreted as the total filling of the surface with active regions. From
unpolarized line profile data S91 observed saturation to occur in the
range $0.5 \leq \log{{\tau}_c{\Omega}_n} \leq 0.75$, corresponding
to the range
$3 \leq R \leq 5$.
Following S91 we plot ${\Delta}V$ against
$\log{R}$, (Figure 1), wherein the
saturation levels of V84 and S91 are indicated and $\tau_c$ is derived
from the empirical function of N84 viz.
\begin{eqnarray}
\log{\tau_c} & = & 1.362-0.166x+0.025{x^2}-5.323{x^3}, \hspace{5mm} x>0 \\
\log{\tau_c} & = & 1.362-0.14x, \hspace{36.0mm}               x<0
\end{eqnarray}
where $x=1-(B-V)$.
This function varies little over the colour range of our
G-K dwarfs and closely follows the similar function of V84
with whom we make direct comparisons.
Theoretical computations by others, e.g. G85
and Rucinski and
Vandenberg (1986) show the same qualitative behaviour
of $\tau_c$ with $B-V$.
\begin{figure*}
  \picplace{10cm}
  \caption{$\Delta V$ against $\log R$}
\end{figure*}
Immediately apparent from figure 1 is the continued increase of
photometric amplitude beyond the limit of chromospheric emission line
saturation (indicated).
{\em{We estimate saturation to
occur when the activity parameter is at least $R \approx 30$ or
$\log{R} \approx 1.5$, a factor 6 to 10 times greater than that
inferred from chromospheric indicators.}}
This apparent dichotomy between the behaviour of chromospheric activity
and spot activity
is highly significant and is discussed further in section 4.

In figure 2 we plot the photometric amplitude (${\Delta}V$) against the
logarithm of the axial rotation period ($P_{rot}$).
The upper limit on the period for saturation is at $P_{rot} \approx 12$hrs
as opposed to $\approx3$
days obtained by V84 for solar-type stars.
Interestingly, all of the late K-type ($B-V > 1.15$) stars in our
sample have ${\Delta}V \leq 0.12$ mag.
This feature is also apparent in figure 3; its possible significance
is discussed in section 4.

\begin{figure*}
  \picplace{10cm}
  \caption{$\Delta V$ against $\log P$}
\end{figure*}
The trends of ${\Delta}V$ with both the rotation period and activity
number are very similar, with
the upper bounds of both scatter plots determined by the same stars.
This implies that the variation of convective turnover time
with spectral type is not a strong factor in the maximum
$\Delta\,V$ variations. This is not surprising given that the convective
turnover time changes by less than half an order of magnitude over the entire
(B-V) range in which we are interested.

The physical significance of the upper bounds of these plots is a somewhat
open question. We discuss this issue further in the next section.

Ca II H+K and Zeeman broadening both indicate that the level of
magnetic surface activity is dependent on spectral type (N84, Saar
1987). This is to be expected if the dynamo generation of magnetic fields
is reliant on both convective motions and rotation. The effect of
convection on the sub-surface fields should scale
with the depth of the convective envelope and hence with
spectral type (Durney and Robinson 1982).

To test these assumptions we plotted ${\Delta}V$
(as a tracer for magnetic surface activity) against colour,
($B-V$), (see figure 3). Immediately apparent is a dearth of high
${\Delta}V$ at both the blue and red end of the plot, although the latter
appears less pronounced due to the paucity of data.

\begin{figure*}
  \picplace{10cm}
  \caption{$\Delta V$ against colour}
\end{figure*}
The low lightcurve amplitude among the bluer stars is likely to be
a selection effect, caused by the age constraint imposed on our
sample. By the age of $\alpha$ Persei (the earliest age of our sample)
no early G-dwarfs are observed to possess high
rotation rates, having, most likely succumbed to magnetic braking.
Figure 1 of Stauffer (1991) and figures 30 and 31 of Prosser (1991)
demonstrate this observationally
when examining only confirmed cluster members.

In contrast, however, the decrease in ${\Delta}V$ towards low masses
cannot be entirely accounted for by selection, and maybe a result of a
physical property of the stars themselves, which is discussed in the
next section.

\section{Discussion}\label{sec:disc}

\subsection{The significance of an upper bound on ${\Delta}V$}

The continued increase of photometric amplitude beyond the limit
corresponding to chromospheric emission line saturation is statistically
highly significant. If one assumes that the measured amplitudes are
subject to gaussian observational errors of dispersion, $\sigma = 0.014$
mag. (resulting from adding in quadrature errors of 0.01 mag. in each of
the two differential photometry measurements), then simple
likelihood analysis applied to Figure 1 reveals that the probability of
measuring the observed amplitudes given the `chromospheric' saturation
level of $R \approx 5$ is less than $10^{-50}$ !. Moreover, this probability
increases to $0.01$ only for $\log R \, \approx \, 1.5$. In other words
there is a $99 \%$ probability that the ${\em true \/}$ saturation level is
higher than $R \approx 30$. Comparable significance levels
are obtained from a similar analysis of Figure 2. It would seem, therefore,
that our claim that the photometric amplitude saturates at a level
considerably higher than chromospheric line emission is statistically very
robust.

It is interesting to consider, however, whether we can obtain a more precise
constraint on the level at which photometric saturation occurs and, more
importantly, establish how closely this level corresponds to the saturation
of the stellar dynamo. If this is to be possible we must first examine the
reasons for the observed scatter in Figures 1 and 2, and consider whether we
are underestimating the {\em intrinsic \/} scatter -- and hence the true
saturation level -- as a result of observational selection effects.

There are a number of factors which account for
a spread in the observed values of ${\Delta}V$ for stars of a given activity
parameter or rotation period. Perhaps the most obvious of these factors
results from
the activity cycle of the observed stars, and in particular the fact that the
measured amplitudes are no more than a `snapshot' of each star at different
points in that activity cycle (c.f. Bailunas and Vaughan, 1985; Innis et al
1988). It is clear from Figure 1 that our sample is too sparse at low values
of the activity parameter $(\log R \leq 1.5)$ to allow a meaningful
assessment of the effects of the activity cycle to be made in those regimes.
For $\log R > 1.5$, however, the number of observed stars in our sample is
considerably larger and covers a narrower colour range. Our sample includes
24 stars with $0.70 < (B-V) < 1.1$, and $1.5 < \log R < 2.0$. Restricting
the colour interval to $0.8 < (B-V) < 1.0$, we still find 16 stars in our
sample with $1.5 < \log R < 2.0$. If we
assume that the amplitudes measured for these stars approximate to a fair
sample of the range of amplitudes which one would observe over the
{\em complete \/} activity cycle of a typical single star in this colour
range, then we might
expect that the maximum observed photometric amplitude would indeed correspond
to a star close to the maximum of its activity cycle. Putting this in simpler
terms, our sample is sufficiently large in the range $\log R > 1.5$ to make it
seem unlikely that {\em all \/} of the stars have been `snapped' at
low points in their activity cycle. Unfortunately accurate observations of
stellar activity cycles remain too limited to allow any more
quantitative analysis of this effect (Innis et al 1988). It nevertheless
seems reasonable to
conclude that we have not significantly underestimated the true saturation of
photometric amplitudes as a result of selection effects related to the stellar
activity cycle.

The other main factor responsible for the spread in the observed
values of ${\Delta}V$ for stars of a given activity parameter or rotation
period is the intrinsic distribution of stellar inclinations.
We can estimate the inclination of each star in our sample by combining its
measured $v \sin i$ with its period and an estimate of its angular diameter
(and hence radius) derived from the magnitude--surface brightness relation
of Barnes, Evans and Moffett (1978), as recalibrated in O'Dell, Hendry and
Collier Cameron (1994). The inclination estimates obtained by this
method are given in Table 1, together with the observed $v \sin i$ of each
star, where measured.

Since the intrinsic scatter in the Barnes--Evans relation -- and consequently
the dispersion of the angular diameter estimate of each star -- is of
the order of $10 \%$, it is difficult to reach any firm conclusions about
the impact of
inclination selection effects on Figures 1 and 2. It is instructive to note,
however, that the mean estimated inclination of stars in our sample with
photometric
amplitudes $\Delta V > 0.1$ mag. is only $ \sim 25 ^{\circ}$. This suggests
that it may be possible to obtain larger photometric amplitudes than those
measured in our sample if one were able to observe stars at more favourable
(i.e. higher) inclinations. In other words the upper bound apparent in
Figures 1 and 2 may be in part an artefact of the inclinations of the selected
stars; consequently any {\em true \/} saturation level may indeed be
somewhat higher than the maximum photometric amplitude which we observe.

It is important to note that, even if our sample {\em were \/} to
provide a clear indication of (as opposed to a lower limit for) saturation of
the photometric amplitude, this would not in general offer conclusive
evidence that dynamo saturation had also occurred. One can envisage several
reasons why an increase in $\Delta V$ might be supressed, while magnetic
activity continued to increase. For example, there exists some evidence that
spot activity is concentrated at higher latitudes in more rapid rotators
(c.f. Strassmeier et al, 1993; Collier Cameron and Unruh, 1994), so that
the change in the cosine of the foreshortening angle for such high latitude
features would reduce the amplitude of spot modulation in these stars.
Secondly, an increase in the total coverage of spots may ultimately
{\em reduce \/}
the overall departure from axisymmetry -- as the greater part of the stellar
disk is filled -- and hence the photometric amplitude. It seems unlikely that
this second effect would be significant in our sample, however: amplitudes as
large as $\Delta V \simeq 0.5$ have been detected in e.g. the sub--giant II
Peg (Avgoloupis et al, 1992; Doyle et al, 1992), suggesting that the
maximum possible spot amplitude is considerably in excess of the levels which
we observe. Notwithstanding this result, none of the dwarf stars considered in
Hall (1991) display V band photometric amplitudes in excess of
$\Delta V \approx 0.3$ mag, so that the question of magnitude suppression
in such stars could still be regarded as open to debate.

In summary, Figures 1 and 2 provide clear evidence
that photometric amplitude -- and hence magnetic activity -- continues to
increase well beyond the level at which chromospheric emission line
indicators are observed to saturate.
It is possible, however, that the upper boundary apparent in both
Figures 1 and 2 may have no physical interpretation, but may be the result
of observational selection effects -- particularly with regard to the
inclinations of the sample stars. In this respect it is worth
noting that the interpretations of N84 and Montesinos and Jordan (1993), of the
chromospheric emission in terms of the dynamo and surface magnetic fields, uses
mean values for the CaII chromospheric emission, over a number of measurements,
for any star, and a mean value for the CaII emission at any (B-V), so that
their analyses assume that the upper bound has no physical significance in
relation to their physical conclusions.

Although the upper boundary to our sample cannot, therefore, be
regarded as conclusive evidence that saturation of photometric amplitude has
occurred, it nevertheless provides a statistically robust lower limit -- for
both photometric amplitude and dynamo saturation -- the physical
implications of which we now discuss in more detail.

\subsection{The implications of dynamo saturation at a higher rotation rate}

If the dynamos of solar-type stars saturate at rotation rates at
least six or more times greater than those implied by
chromospheric indicators, then a number of important implications arise.

The limit of dynamo saturation places severe constraints on magnetic
braking models that have to explain the rapid decrease in rotation rate
for young fast rotating G-dwarfs.
Stauffer (1985) observed that the rotation distribution for solar-type
stars of the $\alpha$ Persei cluster exhibits a strong narrow peak at
slow rotation rates and a tail at rapid rotation rates.
The well known decay law of Skumanich (1972)
($\Omega \, {\propto} \, t^{-1/2}$)
fails to explain why, by the age of the Pleiades (${\sim}70$Myr), this
rapid rotating population has effectively disappeared and only remnants
of the tail exist, consisting of moderate rotators with
$v{\sin}i \leq 45$\kmsec (Stauffer 1991).
Stauffer and Hartmann (1987) addressed this problem by invoking a
core-envelope decoupling scenario whereby only the convective envelope
was braked. Although this model achieved a good fit to the observed
behaviour of young clusters, Li and Collier Cameron (1993) pointed out
that it imposed an unrealistically low upper limit on the magnetic field
strength at the core-envelope interface.
In all braking models dynamo saturation acts to reduce the
power law index of the braking law (Mestel 1988).
CL94 construct, at high rotation rates, a braking model without
core-envelope decoupling, taking into account
thermo-centrifugal wind acceleration and saturated
and unsaturated regimes. Their braking model uses the split monopole
field configuration of Weber and Davis (1967) and in its simplest form is:
\begin{eqnarray}
\dot{\Omega} & = & -k{\Omega}^3 \hspace{27mm} \mbox{unsaturated regime} \\
\dot{\Omega} & = & -k\tilde{{\Omega}}^2{\Omega} \hspace{24.5mm} \mbox{saturated
regime}
\end{eqnarray}
where k is the braking coefficient and $\tilde{\Omega}$ is the rotation
rate at which the dynamo saturates. Over long timescales
the braking model for the unsaturated regime reduces
to the Skumanich decay law.
The braking rate is clearly less efficient in the saturated regime. CL94
therefore argue that for
solar-type stars to be effectively braked within the 20Myr timespan
separating the the age of $\alpha$ Persei and the Pleiades, dynamo
saturation must occur at an inverse Rossby number (or rotation rate)
that is four to five times greater
than the value at which chromospheric saturation sets in: in other words
$\tilde{\Omega}$ must lie somewhere between the rotation
rates of the fast rotating $\alpha$ Persei G-dwarfs and their
Pleiades counterparts.
This onset of saturation is in
close agreement with the results of section 3,
although the minimum level of
saturation inferred from spot coverage appears at a
slightly higher rotation rate.

The braking model of CL94 also suggests that lower mass stars may saturate
at lower rotation rates than the higher mass stars,
in which case they must spend a longer period of time spinning down in
the saturated state.
Following the evolution of angular momentum from the $\alpha$
Persei to Hyades clusters, the rapidly rotating K-dwarfs are indeed
observed to spin down at a later epoch than the G-dwarfs (Stauffer 1991).
Examining figures 2 and 3 we notice that all of the late K stars exhibit
low photometric amplitudes, possibly indicating that the lower
mass stars attain saturation at a slower rotation rate.
It is unlikely that this is due entirely to a change in the
temperature difference and hence contrast
between the photosphere and
spots. The factor that will determine a significant change in
${\Delta}V$ is the ratio of these temperatures ($T_{spot}/T_{phot}$)
and this varies little
over the spectral range of our sample.
Alternatively the deeper convective envelope of the lower mass stars may
allow the magnetic flux to reach higher latitudes at high rotation
rates.
This would act to increase spot coverage
at high latitudes which in turn would act to reduce ${\Delta}V$.
We consider this result inconclusive at present.

The mismatch between the chromospheric level of saturation and that
obtained from starspot coverage has other physical implications. Given
that the disc continues to fill with surface features beyond the limit of
chromospheric saturation, this must
question whether the chromospheric indicators are a reliable diagnostic
of the dynamo process.
On the basis of this result, and the fact that
current magnetic braking theory demands dynamo
saturation to occur at considerably higher rotation rates, we propose
that
spot coverage acts as a better diagnostic for dynamo saturation at
rapid rotation rates. Methods employing chromospheric
and transition region
indicators on solar-type stars work well when the spotted regions are
so small as to be undetectable.

\section{Summary}\label{sec:summ}

In this paper we have presented statistically robust evidence, from the
photometric observations of young cluster and field G and K-type stars,
indicating that saturation of starspot activity
in solar-type stars occurs at a rotation rate at least
6 - 10 times greater than that found from chromospheric emission
and unpolarized Zeeman-broadened photospheric line profiles.
This result is consistent with the recent requirements of magnetic
braking models that can reproduce the evolution of angular momentum in
solar-type stars from the early main-sequence to the age of the Sun
without invoking core-envelope decoupling.

\section{Acknowledgements}

MAO wishes to thank Dr Charles Prosser for a compilation of all the known
Pleiades and $\alpha$ Persei period - amplitude data.
The authors acknowledge use of the Microvax 3400 computer, funded by a
SERC grant to the Astronomy Centre at the
University of Sussex. MAO was funded by an SERC research studentship,
PMP and MAH were supported by SERC research fellowships and ACC was
suported by an SERC advanced fellowship at the University of Sussex.
\\
\\
\\
\newpage
\noindent
\,\,\,{\bf References}
\\
\begin{description}
\item[]Avgoloupis A., Doyle J.G., Mavridis L.N., Seiradakis J.H.,
Mathioudakis M., 1992. In: Byrne P.B., Mullan D.J. (eds.) Surface
Inhomogeneities in Late--Type Stars. Springer--Verlag, p247
\item[]Bailunas S.L., Vaughan A.H., 1985, {\em ARA\&A}, {\bf 23}, 379
\item[]Barnes T.G., Evans D.S., Moffett T.J., 1978, {\em MNRAS}, {\bf 183},
298
\item[]Collier Cameron A., Li J., 1994, {\em MNRAS}, submitted (CL94)
\item[]Collier Cameron A., Unruh, Y.C., 1994, {\em MNRAS}, submitted
\item[]Dorren J.D., Guinan E.F., Dewarf L.E., 1993, personal
communication
\item[]Doyle J.G., Bromage G.E., Collier Cameron A., Kilkenny D.W.,
Krzesinski J., Murphy H.M., Neff J.E., Pajdosz G., van Wyk F., 1992.
In: Byrne P.B., Mullan D.J. (eds.) Surface Inhomogeneities in Late--Type
Stars. Springer, Berlin, p276
\item[]Durney B.R., Robinson R.D., 1982, {\em ApJ}, {\bf 253}, 290
\item[]Gilliland R.L., 1985, {\em ApJ}, {\bf 299}, 286
\item[]Hall D.S., 1991. In: Tuominen I., Moss D., R\"{u}diger G.
(eds.) The Sun and Cool Stars: Activity, Magnetism, Dynamos. Springer,
Berlin, p.353
\item[]Innis J.L., Thompson K., Coates D.W., Lloyd Evans T., 1988,
{\em MNRAS}, {\bf 235}, 1411
\item[]Innis J.L., Coates D.W., Thompson K., Lloyd Evans T., 1990,
{\em MNRAS}, {\bf 242}, 306
\item[]Kraft R.P., 1967, {\em ApJ}, {\bf 150}, 551
\item[]Kron G.E., 1952, {\em ApJ}, {\bf 115}, 301
\item[]Li J., Collier Cameron A., 1993, {\em MNRAS}, {\bf 261}, 766
\item[]Linsky J., 1990. In: Ulmschneider P., Priest E., Rosner R. (eds.)
Mechanisms of Chromospheric and Coronal Heating. Springer, Berlin, p.166
\item[]Magnitskii A.K., {\em SvA Lett}, {\bf 13}, 451
\item[]Mermilliod J-C., 1981, {\em A\&A}, {\bf 97}, 235
\item[]Mestel L., 1988. In: Havnes O. et al. (eds.) Activity in Cool
Star Envelopes. Kluwer Academic Publishers, Dordrecht, p.1
\item[]Montesinos B., Jordan C., 1993, {\em MNRAS}, {\bf 264}, 900
\item[]Noyes R.W., 1983. In: Stenflo J.O. (ed.) Solar and Stellar
Magnetic Fields: Origins and Coronal Effects. Dordrecht:Reidel, p.133
\item[]Noyes R.W., Hartmann L.W., Baliunas S.L., Duncan D.K., Vaughan
A.H., 1984, {\em ApJ}, {\bf 279}, 763 (N84)
\item[]O'Dell M.A., Collier Cameron A., 1993, {\em MNRAS}, {\bf 262}, 521
\item[]O'Dell M.A., Collier Cameron A., Hilditch R.W., Bell S.A.,
1994, {\em MNRAS}, in preparation
\item[]O'Dell M.A., Hendry M.A., Collier Cameron A., 1994, {\em MNRAS},
in press
\item[]Parker E.N., 1955, {\em ApJ}, {\bf 121}, 491
\item[]Prosser C.F., 1991, Ph.D. thesis, Univ. California.
\item[]Prosser C.F., Schild R.E., Stauffer J.R., Jones B.F., 1993a, {\em
PASP}, {\bf 105}, 269
\item[]Prosser C.F., Shetrone M.D., Marilli E., Catalano S., Williams
S.D., Backman D.E., Laaksonen B.D., Adige V., Marschall L.A., Stauffer
J.R., 1993b, {\em PASP}, {\bf 105}, 1407
\item[]Rucinski S.M., 1983, {\em A\&AS}, {\bf 52}, 281
\item[]Rucinski S.M., Vandenberg D.A., 1986, {\em PASP}, {\bf 98}, 669
\item[]Saar S., 1988, {\em ApJ}, {\bf 324}, 441
\item[]Saar S., 1991. In: Tuominen I., Moss D., R\"{u}diger G.
(eds.) The Sun and Cool Stars: Activity, Magnetism, Dynamos. Springer,
Berlin, p.389 (S91)
\item[]Skumanich A., 1972, {\em ApJ}, {\bf 171}, 565
\item[]Stauffer J.R., 1991. In: Catalano S., Stauffer J.R. (eds.)
Angular Momentum
Evolution of Young Stars. Kluwer Academic Publishers, Dordrecht, p.117
\item[]Stauffer J.R., Hartmann L.W., Burnham J.N., 1985, {\em ApJ},
{\bf 289}, 247
\item[]Stauffer J.R., Hartmann L.W., 1987a, {\em ApJ}, {\bf 318}, 337
\item[]Stauffer J.R., Schild R.E., Baliunas S.L., Africano J.L., 1987b,
{\em PASP}, {\bf 99}, 471
\item[]Stauffer J.R., Hartmann L.W., 1989, {\em ApJ}, {\bf 346}, 160
\item[]Strassmeier K.G., Rice J.B., Wehlau W.H., Hill G.M., Matthews, J.M.,
1993 {\em A\&AS}, {\bf 268}, 671
\item[]Van Leeuwen F., Alphenaar P., 1982, {\em ESO messenger}, {\bf
28}, 15
\item[]Vilhu O., 1984, {\em A\&A}, {\bf 133}, 117 (V84)
\item[]Vilhu O., Rucinski S.M., 1983, {\em A\&A}, {\bf 127}, 5
\item[]Weber E.J., Davis L., 1967, {\em ApJ}, {\bf 148}, 217

\end{description}

\end{document}